# Polarization dependent interface properties of ferroelectric Schottky barriers studied by soft X-ray absorption spectroscopy

H. Kohlstedt*, A. Petraru, and M. Meier

*Institut für Festkörperforschung and CNI, Forschungszentrum Jülich,*

*D-52425 Jülich, Germany*

J. Denlinger, J. Guo, Y. Wanli, A. Scholl, and B. Freelon

*Advanced Light Source, Lawrence Berkeley National Laboratory,*

*Berkeley California 94720, USA*

T. Schneller and R. Waser

*Institute of Materials for Electronic Engineering 2, RWTH-Aachen,*

*Sommerfeldstr. 24, D-52074 Aachen, Germany*

P. Yu and R. Ramesh

*Department of Materials Science and Engineering and Department of Physics,*

*University of California, Berkeley, California 94720 USA*

T. Learmonth, P.-A. Glans and K. E. Smith

*Department of Physics, Boston University, Boston, Massachusetts 02215 USA*

We applied soft X-ray absorption spectroscopy to study the Ti $L$-edge in ferroelectric capacitors using a modified total electron yield method. The inner photo currents and the X-ray absorption spectra were polarization state dependent. The results are explained on the basis of photo electric effects and the inner potential in the ferroelectric capacitors as a result of back-to-back Schottky barriers superimposed by the potential due to the depolarization field. In general, the presented method offers the opportunity to investigate the electronic structure of buried metal-insulator and metal-semiconductor interfaces in thin film devices.

*Corresponding author: h.h.kohlstedt@fz-juelich.de*

Interface characteristics are of crucial importance in novel electronic devices. Field effect transistors, tunnel junctions and laser diodes are only a few examples of a hardly manageable number of electronic devices in which interfaces are essential.[1] Electronic devices realize their full potential when an external electric field is applied. The application of an electric field to devices, being simultaneously subject to photon irradiation, is an interesting approach to study a rather broad spectrum of fundamental issues and practical device oriented questions. By using ultraviolet (UV) light, internal photo current measurements were successfully applied to metal-oxide-semiconductor (MOS) structures under bias. This allowed the evaluation of the energetic and spatial distribution of trapped charges in the oxide and the effective barrier height for example.[2-5] Recently this technique was extended to MOS capacitors under low bias fields and to high-$k$ dielectrics. [6,7] Kobayashi and coworkers investigated Pt/oxide/p-InP (100) MIS Schottky diodes under an external applied bias potential whilst simultaneously taking x-ray photoelectron spectra (XPS).[8] M. Ishii et al. developed a capacitance X-ray absorption fine structure (XAFS) method to analyze site selective electron trapping centers in $Cu_2O$ and $GaO$.[9] Hard x-ray diffraction has been applied to evaluate the structural changes in piezoelectric and ferroelectric materials under an electric bias field, with charge-density variations due to the applied electric field being detected.[10] Recently such studies have focused on the microscopic definition of polarization.[11] Janousch et al. employed X-ray near edge spectroscopy (XANES) to investigate the local environment of Cr in resistive switching elements using finger electrodes.[12] The mechanism of field-effect doping in cuprate superconductors was studied using soft X-ray absorption spectroscopy and an applied gate voltage.[13] Motivated by the fact, that soft X-ray absorption spectroscopy (SXAS) is a unique tool to study the electronic structure in metals, semiconductors and insulators, we applied this technique to investigate the buried ferroelectric-electrode interface of ferroelectric capacitors.[14-16] SXAS measures the transition probabilities between a core level and the unoccupied electronic states of a material at an atomic site.[17,18] The total fluorescence yield (TFY) monitors bulk properties due to the large photon attenuation length (100 nm –200 nm), while the short mean free path of photoelectrons (< 3 nm) coming from the sample surface makes the total electron yield (TEY) a surface sensitive technique.[17,19-21]

The Pt/PbZr$_{0.3}$Ti$_{0.7}$O$_3$ (PZT)/Pt ferroelectric capacitors were deposited on platinized Si/SiO$_2$ substrates, which results in a whole layer stack (from substrate to top electrode) of Si(600µm)/SiO$_2$(450nm)/Ti(15nm)/Pt(150nm)/PZT(180nm)/Pt(50nm). Chemical solution deposition was utilized to grow the PZT film[22] and the capacitor areas were 3.425 mm$^2$. Fig. 1 (a) shows a ferroelectric hysteresis loop of such a device after an annealing at 700°C for 3 min in oxygen. For the as fabricated capacitors (black curve) the coercive fields +/- E$_c$ were determined to be +/-110 kV/cm (V$_c$ = +/-2V). The remnant states were +/-$P_r$ = 30 µC/cm$^2$, which compares well with the literature (black loop 1 in Fig. 1(a).[22] Such ferroelectric capacitors were placed in the undulator beam line 8.0 at the Advanced Light Source of the Lawrence Berkeley Laboratory.[23] About 80% of the capacitor area was illuminated by the synchrotron beam. The top Pt layer acts as a semi-transparent electrode for the soft X-rays. For the SXAS experiments the two leads of a capacitor were connected to an external circuit, which is sketched in Fig. 1 (b). The two ammeters connected to ground allowed the study of the current distribution under soft X-ray radiation, whilst a typical TEY standard set-up uses only the bottom ammeter to study the surface properties of the bare surface without a top electrode.[17] During the entire experiment the capacitor was under short circuit boundary conditions. The capacitor was initially biased with a voltage pulse for 1 sec, either of +4 V or –4 V, between the top electrode and the bottom electrode (+4 V means positive 4 V poling on the top electrode and vice versa for a –4V pulse setting) to obtain a mono domain state. While tuning the energy of incoming soft X-rays between 450 eV and 480 eV, the current in the top ($I_{top}$) and in the bottom ($I_{bot}$) leads were simultaneously acquired by the ammeters. We observed four prominent peaks (see spectrum $I_{bot}$ (–$P_r$) in Fig. 2) which are related to the crystal field splitting of Ti$^{4+}$ in an octahedral symmetry. The first two peaks (at 458 eV and 460 eV) belong to the $2p_{3/2}$ ($t_{2g}$ and $e_g$) ($L_3$ edge) and the third and fourth peak (464.35 eV and 466.7 eV) belong to the $2p_{1/2}$ ($t_{2g}$ and $e_g$) ($L_2$ edge) transitions.[24,25]

The spectra shown in Fig. 2 (a), exhibit a number of interesting features. We observed four different base line currents (a base line current is defined at the photon energy of 450 eV). Obviously, these currents ($I_{top}$ and $I_{bot}$) depend on the polarization state (+$P_r$ and -$P_r$). The current $I_{bot}$ (Fig. 2(a) showed the sign of a standard TEY set-up, i.e. the SXAS exhibited peaks, independent on the ferroelectric capacitors polarization state, whereas the Ti $L_{3,2}$ edge absorption signal $I_{top}$ exhibited dips for the +$P_r$ and the -$P_r$ polarization state (curves labeled by $I_{top}$ (+/-$P_r$), $I_{bot}$ (+/-$P_r$) in Fig. 2 (a)). Moreover $\Delta I_{top}$ (we used the current of the $t_{2g}$ Ti $L_3$ edge at 458 eV) in the +$P_r$ state was roughly a factor 2 smaller than $\Delta I_{top}$ in the –$P_r$ state. In addition, the spectra in the –$P_r$ state trace the well known peak sequence of the Ti $L$ edge in PZT,[26, 27] while the spectra in the +$P_r$ state exhibited a broadening of the Ti $L_3$ $e_g$ peak and a line splitting of the Ti $L_2$ $e_g$ peak at ~ 466 eV.

A possible scenario could be the following: Photo electrons were excited by the soft X-rays at the Pt (50 nm thick) top electrode surface. The electrons have sufficient kinetic energy to leave the top Pt film into vacuum (external photo electric effect) and therefore the capacitor (the top Pt electrode) is positively charged (see Fig. 1 (b), process 1). Consequently, a neutralization current $I_{neut}$ from ground to the capacitor is generated for compensation (see Fig. 1(b)). $I_{neut}$ splits into contributions to $I_{top}$ and $I_{bot}$.

For the $+P_r$ state for example, $I_{top}$ = 19.8 nA (@450 eV), $I_{bot}$ = 1.1 nA (@450 nA) and $I_{neut} = I_{top} + I_{bot}$ = 20.9 nA. For the $-P_r$ state these currents at 450 eV are $I_{top}$ = 12 nA and $I_{bot}$ = 8.9 nA and therefore $I_{neut}$ = 20.9 nA. In the energy range between 400 eV and 600 eV no core level excitations of Pt exist. This results in a constant $I_{neut}$ in this photon energy range. The average current $I_{ave} = (I_{top} + I_{bot})/2$ is about 10 nA for each polarization state (see Fig. 2 (a)). Further contributions to $I_{top}$ and $I_{bot}$ originate from photo carriers excited at the Ti $L$-edge thresholds between 450 and 480 eV (internal photo effect) in the PZT. Two major internal photo carrier effects might be relevant. First, the photon excitation energy is more than two orders of magnitude larger than the band gap of the ferroelectric PZT, which is of about 3.4 eV.[16] Therefore electron-hole pairs were generated in the PZT (see Fig. 1 (b), process 2). Second, in case the photon energy reached the Ti $L_{2,3}$-edges, additional electron carriers are generated due to core level excitation processes. In contrast to the first internal photo effect, the core holes are strongly localized at atomic sites and therefore cannot move under an electric field (see Fig. 1 (b), process 3). In light of the 3 photo excitation processes, we can qualitatively explain the results shown in Fig. 2 (a), if we take into account the inner potential distribution (for $+P_r$ and $-P_r$) and the inhomogeneous photon flux inside the PZT. In Fig. 2 (b) the charge distribution is schematically shown for the $+P_r$ state. The conduction and valence energy bands for $+P_r$ (red curves) and $-P_r$ (blue curves) are sketched in Fig. 2 (c) for short circuit boundary conditions. Here we assume a (slight) p-type character of the PZT bulk with the tendency to an n-type semiconductor at both interfaces (see black lines for $E_v$ (x) and $E_c$ (x) for a hypothetical, paraelectric state of the PZT). For simplification, the band diagrams are pinned at both interfaces (e.g. due to defects). We are aware of the fact, that the band diagram of the Pt/PZT interface is still under debate and the exact band diagrams may differ correspondingly.[16] The bands of our back-to-back Pt/PZT/Pt Schottky barrier device are superimposed on the potential which originates from the depolarization field (inside the ferroelectric).[32] The later is a consequence of the non-perfect screening (not shown here) of the ferroelectric bound charge at the metal-insulator interfaces of a ferroelectric capacitor.[28] Therefore we expect a photo current $I_{photo}$ (from the internal photo effect) in the external circuit with an opposite sign which reduces $I_{bot}$ and adds to $I_{top}$. This current has no influence on the neutralization current $I_{neut} = I_{bot} - I_{photo} + I_{top} + I_{photo}$. Please note that $I_{neut}$ is constant, while it's splitting into $I_{top}$ and $I_{bot}$ depend on the number of internal photo carriers generated in the PZT. Switching the capacitor from the $+P_r$ state to the $-P_r$ state, the potential due to the depolarization field is reversed (see Fig. 2 (c)). It is likely that the inner potential distribution has a strong influence on the photo carrier transport current, which would explain the distinguish base line currents for the $+P_r$ and $-P_r$ state.

Another important factor is the exponential decay of the photon flux within the ferroelectric. In Fig. 2 (d) the photon flux attenuation described by the law of Lambert-Beer is schematically shown for two photon energies. At the upper Pt/PZT interface the flux is the same but the attenuation coefficients depend on the photon energy. Two curves, at 450 eV (pre-edge region) and at 458 eV (at the Ti$L_3$ $t_{2g}$ edge) are qualitatively compared. Overall the photo current of the irradiated ferroelectric capacitor is determined by the band diagram of metal-insulator-metal, back-to back Schottky photo diodes [29-31] with the additional degree of freedom from the switchable polarization weighted by the photon flux decay inside the PZT. The sensitivity of photo currents on the polarization state was observed by I. Boerasu et al. using UV light at 340 nm.[32] We like to point out that the use of soft X-rays delivers simultaneously interesting additional information of the electronic interface structure.

To clarify the different form of the SXAS signal at the Ti $L_{3,2}$ edge for the $+P_r$ and $-P_r$ state, we compared both signals in Fig. 3 (taken the data $I_{bot}$ (+/-$P_r$)). The signal of the $+P_r$ state (of the Ti $L_3$ edge, $t_{2g}$ peak) from of $I_{top}(+Pr)$ was enhanced to achieve the same amplitude as the Ti $L_3$ $t_{2g}$ peak of the $–P_r$ state. The considerable deviation found at the $L_2$ $e_g$ peaks (double peak structure in the $–P_r$ state, red curve) indicate a rather different local environment of Ti atoms at the top PZT/Pt interface in dependency of the polarization state. We observed a crystal field splitting $\Delta_{CF}$ of about 2 eV as indicated in Fig. 3. A closer look to the crystal field splitting in the $P_r$ and $–P_r$ states showed a clear difference in the crystal field splitting if between both polarization states. In the inset in Fig. 3 the $I_{top}$ $(+/-P_r)$ are compared and we found a difference $\Delta^*_{CF}$ = 0.29 eV. It is unlikely that changes in the Ti crystal field are due to ferroelectric switching process alone. The extremely small structural changes between the $–P_r$ and $+P_r$ states might *not* be relevant for such drastic changes in the observed Ti absorption signals. We suggest here another scenario. The different spectra shown in Fig. 3 became more significant throughout the experiment and might be related to the high photon flux of the synchrotron radiation. Indeed in $TiO_2$ photo catalytic effects are known to be relevant.[33] In our case the strong depolarization field inside the capacitors may lead to a considerable diffusion of oxygen under the electric field and the irradiation. We unintentionally crossed the border from a non-destructive to a (partly) destructive experiment. This idea is supported by the fact that after the 10 hours of beam line experiments the PZT was still ferroelectric but a rather drastic change of the *P* vs. *E* hysteresis loop occurred (see red loop Fig. 1 (a)). This effect was even more pronounced after experiments at the oxygen *K*-edge (see blue loop in Fig. 1(a)). On the other hand, the asymmetry in the *P* vs. *V* loop (blue curve in Fig. 1(b) and a possible relation to the absorption signal raised an interesting speculation. Is it possible that the absorption signals $I_{top}$ $(+P_r)$ and $I_{top}$ $(-P_r)$, as shown in Fig. 3, reflect the different electronic structure of the bottom and top PZT/Pt interfaces? Indeed, the leakage behavior of the ferroelectric capacitor (blue curve in Fig. 1(b)) for the positive and negative bias, indicate a different electronic structure between the top and bottom interface. From this perspective it is reasonable to assume that the absorption signals probe either the electronic structure of the top or bottom interface in dependency on the polarization state. The strongly distorted signal in the $+P_r$ state means that this signal might be related to the top interface. First because the photon flux is much larger (Fig. 2 (d)) at the top interface and so photo catalytic effects are more likely than at the bottom interface and because the leakage behavior is worse for the positive bias which could be related to a disturbed Ti environment (for example oxygen deficiency). Nonetheless, because the exact inner potential distribution in the PZT is unknown, we cannot role out that the spectra reflect to some extend the difference of the photo current between both interfaces but leaved a net photo current from the bottom to the top electrode. Further experiments are necessary and the additional application of small bias field might be helpful to clarify this issue.

To further support the distinguish role of the depolarization field, we investigated non-ferroelectric Pt/SrTiO$_3$/Pt capacitors. Beside the dielectric material all other parameters (thicknesses, deposition method etc.) were the same as for the Pt/PZT/Pt capacitor. Indeed we could not find any indication of ferroelectricity in the $P$ vs. $V$ loop. The $P$ vs. $V$ loop reflected the properties of a linear, highly isolating dielectric. In Fig. 4 the soft X-ray absorption signals are shown for a +4 V and –4V setting. In contrast to the absorption signals of the ferroelectric capacitor shown in Fig. 2(a), $I_{bot}$ (-$P_r$) and $I_{top}$(+$P_r$) we observed a much less strong shift of the TEY in the Pt/SrTiO$_3$/Pt capacitor. The photo currents ($I_{bot}$ and $I_{top}$) were independent whether the Pt/SrTiO$_3$/Pt capacitor was initially poled by +4V or –4V. This result clearly shows that no depolarization field exists in the non-polar SrTiO$_3$ dielectric. The inset in Fig. 4 shows a linear $P$ vs. $V$ dependency as expected for the linear (non-polar) dielectric SrTiO$_3$. In addition the SrTiO$_3$ experiment gives further evidence that indeed the depolarization field in PZT has a considerable effect on the inner potential distribution of Pt/PZT/Pt capacitors and originates the four distinguish base line currents shown in Fig. 2 (a).

Finally we discuss the probe depth of our experiment. It is known from TEY investigations of the bare surface of insulators that the probe depth is approximately 5 nm at photon energies of several hundred eV. This is a factor $\approx 2$ larger than the probe depth of photoelectrons at pure metal surfaces. This fact reflects the larger cross-section of electrons in metals than in insulators.[17] There is no obvious reason that the probe depth at metal/insulator interfaces is different. Thereofore we assume that we probe the electronic structure up to 5 nm adjacent to the Pt/PZT interfaces if the absorption signal at the Ti $L$-edge of the insulator is measured. Consequently, the method offers an interesting approach to study the electronic structure of the top and bottom dead interfacial layers while the device is fully functioning.

In conclusion we presented a modified total electron yield set-up to study the current distribution in ferroelectric and dielectric capacitors under soft X-ray synchrotron irradiation. We observed a clear difference in the photo currents between ferroelectric Pt/PbZr$_{0.3}$Ti$_{0.7}$O$_3$/Pt and dielectric Pt/SrTiO$_3$/Pt capacitors. Our findings are explained on the basis of the external and internal photoelectric effect, the depolarization field and the Schottky barriers at the top and bottom interface of the capacitors. In addition it turned out, that care is needed with this approach to ensure the experiment to remain non-destructive. This experimental study demonstrated SXAS to be a potential probe for studying the interfacial electronic structure of planar devices under soft X-ray irradiation.

**Acknowledgement**

The author thanks Michael Hambe for carefully reading the manuscript. We thank Terrence Jach for helpful discussions. The Advanced Light Source is supported by the U.S. Department of Energy under Contract No. DE-AC02-05CH11231. The Boston University program is supported in part by the Department of Energy under DE-FG02-98ER45680. The work was supported by the Material Worlds Network (DFG and NSF).

**Figure Captions:**

Fig. 1. (a) Hysteresis loops of a ferroelectric capacitor, for the as fabricated sample (solid black line), after 10 hours of irradiation at the Ti edge (red line) and after additional 2 hours at the O $K$-edge (blue line). (b) Schematic of the modified total electron yield set-up using two ammeters, one in the top and another in the bottom wiring line. Essential photo electric effects are labeled by 1- 3 (b).

Fig. 2. (a) Four soft x-ray absorption spectra $I_{bot}$ and $I_{top}$ observed for the $+P_r$ and $-P_r$ states for a Pt/PZT/Pt ferroelectric capacitor. The four base line currents at 450 eV are labeled by ①-④. (b) Charge distributions in the capacitor for the $+P_r$ state. (c) Energy diagrams ($E_v$ valance band, $E_c$ conduction band) for the PZT in the paraelectric state (black) and the $+P_r$ (red) and $-P_r$ (blue) polarization states. (c) Attenuation of the photon flux inside the PZT for two photon energies.

Fig. 3. Comparison of the SXAS in the $+P_r$ (black) and $-P_r$ state (red) detected by the bottom Ammeter. The signal of the $-P_r$ state was adjusted to the amplitude of the $+P_r$ state for the $L_3$ Ti $e_g$ state (peak at 458 eV). The inset shows a similar comparison for the top ammeter, which exhibited a difference in the Ti-crystal field at the $L_3$ edge.

Fig. 4 Four soft x-ray absorption spectra $I_{bot}$ and $I_{top}$ observed for a +4 V and –4 V settings of a Pt/SrTiO$_3$/Pt dielectric capacitor. The bias voltage $V$ was 0 Volt. Please note that the data acquisition system did not allow detecting both current polarities. Therefore parts of the lower to spectra were cut. This experimental restriction has no influence on the presented interpretation. The inset shows the polarization and displacement current vs. the applied bias voltage for the capacitor.

**References**

[1]  "Nanoelectronics and Information Technology", Wiley-VCH, Weinheim 2003, ed. by R.Waser, Markus Grundmann "The Physics of Semiconductors", Springer-Verlag, Berlin 2006.

[2]  K. W. Shepard, J. Appl. Phys. **36**, 796 (1965).

[3]  R. J. Powell and C. N. Berglund, J. Appl. Phys. **42**, 4390 (1971).

[4]  J. DiMaria, J. Appl. Phys. **47**, 4073 (1976).

[5]  H. Nicollian and J. R. Brews, MOS (Metal Oxide Semiconductor) Physics and Technology (Wiley, New York, 1982).

[6]  M. Prewlocki, J. Appl. Phys. **85**, 6610 (1999).

[7]  D. Feinhofer, E. P. Gusev, and D. A. Buchanan, J. Appl. Phys. **103**, 054101 (2008).

[8]  H. Kobayashi, T. Mori, K. Namba, and Y. Nakato, Solid State Comm. **92**, 249 (1994); H. Kobayashi, T. Mori, T. Mori, Y. Nakato, K. H. Park, Y. Nishioka, Surf. Sci. **326**, 124 (1995).

[9]  M. Ishii, A. Nakao, and T. Uchihashi, Phys. Scripta T **115**, 97 (2005).

[10] I. Fujimoto, Acta Cryst. A **38**, 337 (1982). U. Pietsch, J. Mahlberg, and K. Unger, Phys. Status Solidi B **131**, 67 (1985).

[11] J. Stahn, U. Pietsch, P. Blaha and K. Schwarz, Phys. Rev. B **63**, 165205 (2001).

[12] M. Janousch, G. I. Meijer, U. Staub, B. Delley, S. F. Karg, B. P. Andreasson, Adv. Mat. **19**, 2232 (2007).

[13] M. Sulluzzo, G. Gringhelli, J. C. Cezar, N. B. Brookes, G. M. De Luca, F. Fracassi, and R. Vagilo, Phys. Rev. Lett. **100**, 056810 (2008).

[14] J. F. Scott, C. A. P. d. Araujo, Science, **246** 1400 (1989).

[15] H. Kohlstedt, Y. Mustafa, A. Gerber, A. Petraru, M. Fitsilis,

R. Meyer, U. Böttger, and R. Waser, Microelectronic Engineering **80**, 296 (2005).

[16] J. F. Scott, *Ferroelectric Memories*, Springer Series in Advanced

Microelectronics, (Springer-Verlag, Berlin Heidelberg, New York 2000).

[17] J. Stöhr, NEXFAS Spectroscopy, Berlin, Springer 1996,

Springer Series in Surface Science 25.

[18] F. M. F. de Groot, M. Grioni, J. C. Fuggle, J. Ghijsen, G. A. Sawatzky, and

H. Petersen, Phys. Rev. B **40** 5715 (1989).

[19] A. P. Likirskii and I. A. Brytov, Sov. Phys. Solid State **6**, 33 (1964).

[20] J. Stöhr, J. Vac. Sci. Technol. **16**, 37 (1979).

[21] J. Guo, Int. J. Nanotechnol. **1** 193 (2004).

[22] T. Schneller, R. Waser, J. Sol-Gel Sci. Techn. **42**, 337 (2007) and

R. W. Schwartz, T. Schneller, R. Waser, C. R. Chim. **7**, 433 (2004).

[23] J. J. Jia, T. A. Callcott, J. Yurkas, A. W. Ellis, F. J. Himpsel, M. G. Samant,

J. Stöhr, D. L. Ederer, J. A. Carlisle, E. A. Hudson, L. J. Terminello, D. K. Shuh,

R. C. C. Perera, Rev. Sci. Instrum. **66**, 1394 (1995).

[24] L. Soriano, M. Abbate, A. Fernández, A. R. González-Elipe, J. M. Sanz, Surf.

and Interface Anal. **25**, 804 (1997).

[25] F. M. F. de Groot, J. C. Fuggle, B. T. Thole, G. A. Sawatzky,

Phys. Rev. B **41**, 928 (1990).

[26] V. R. Mastelaro, P.P. Neves, S. R. de Lazaro, E. Longo, A. Michalowicz, and

J. A. Eiras, J. Appl. Phys. **99**, 044104 (2006).

[27] T. Higuchi, T. Tsukamoto, T. Hattori, Y. Honda, S. Yokohama, and

H. Funakubo, Jap. J. Appl. Phys. **44**, 6923 (2005).

[28] I. P. Batra and B. D. Silbermann, Sol. Stat. Comm. **11**, 291 (1972).

[29] S. M. Sze, D. J. Coleman, and A. Loya, Solid-St. Electron. **14**, 1209 (1971).

[30] M. Grundmann, "The Physics of Semiconductors" Springer Berlin, Heidelberg, New York 2006.

[31] V. M. Fridkin, *"Ferroelectric Semiconductors"*, Plenum Publishing Corporation, 1980.

[32] I. Boersasu, L. Pintilie, M. Pereira, M. I. Vasilevsky, and M. J. M. Gomes, J. Appl. Phys**. 93**, 4776 (2003).

[33] Amy J. Lisebigler, G. Lu, J. T. Yater, Chem. Rev. **95**, 735 (1995).

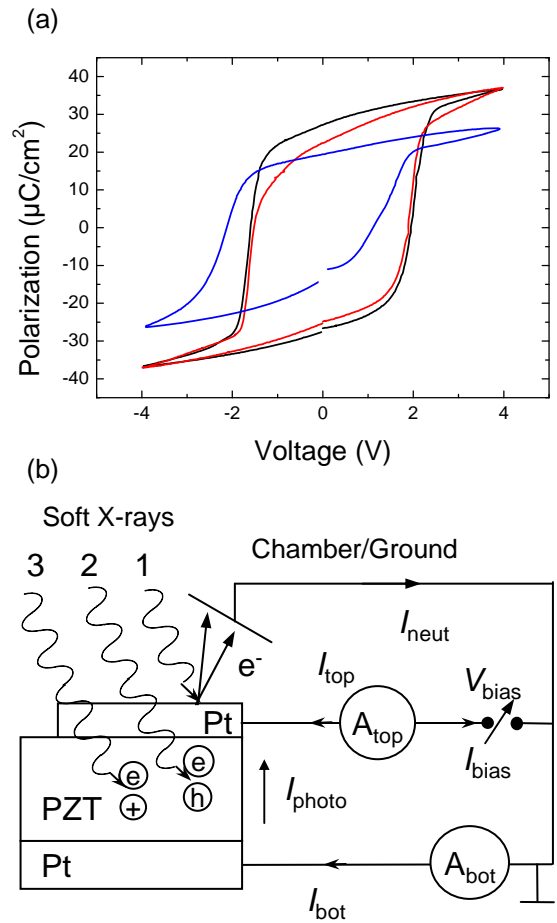

Fig. 1 H. Kohlstedt et al., Polarization dependent interface properties ...

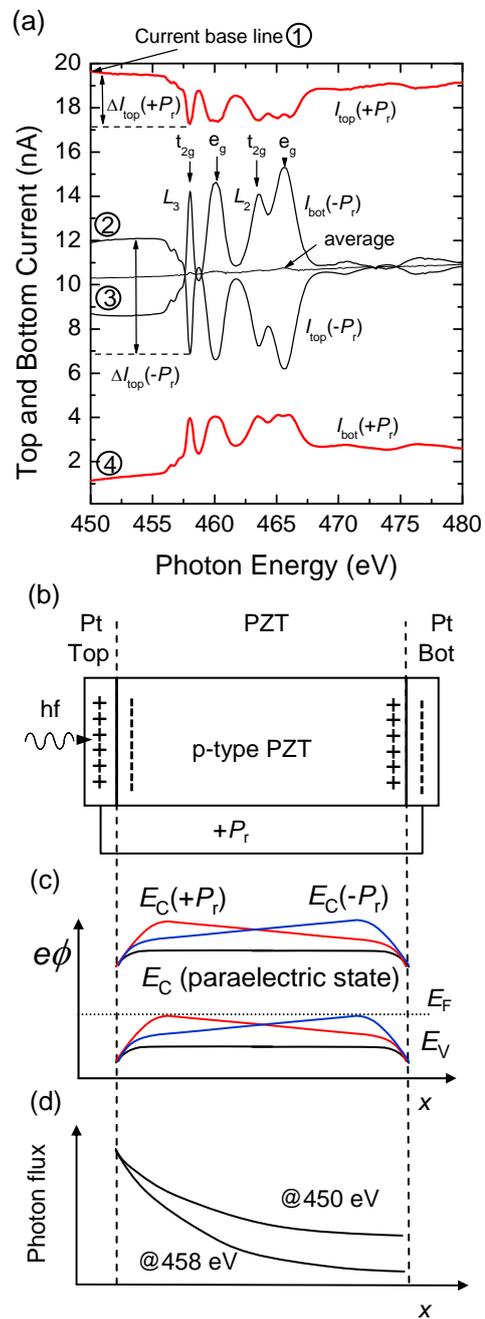

Fig. 2 H. Kohlstedt et al., Polarization dependent interface properties ...

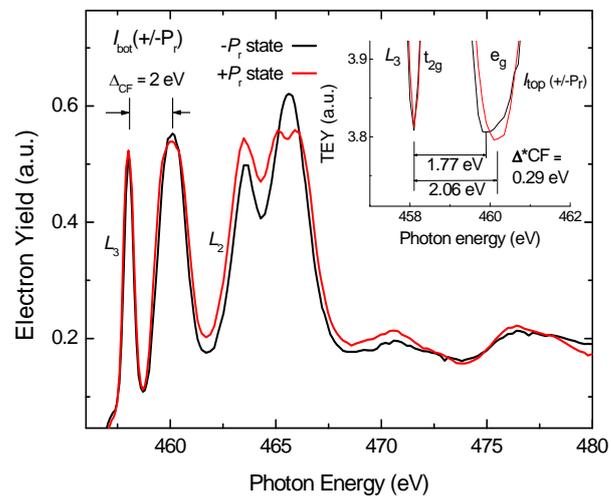

Fig. 3   H. Kohlstedt et al., Polarization dependent interface properties ...

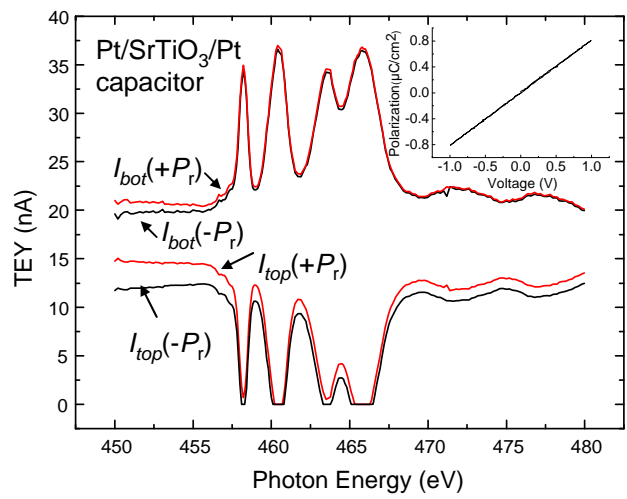

Fig. 4  H. Kohlstedt et al., Polarization dependent interface properties ...